\documentstyle[aps,twocolumn]{revtex}
\begin{document}
\preprint{USACH-FM-00-06}
\title{Off-Shell Effective Lagrangian for NRQCD and Heavy Quarks
Effective Theory}
\author{ J. Gamboa$^1$\thanks{E-mail: jgamboa@lauca.usach.cl}, S.
Lepe$^{1,2}$ \thanks{E-mail: slepe@lauca.usach.cl,} and
L. Vergara$^1$\thanks{E-mail: lvergara@lauca.usach.cl}}
\address{$^1$Departamento de F\'{\i}sica, Universidad de Santiago de Chile,
Casilla 307, Santiago 2, Chile\\
$^2$ Instituto de F\'{\i}sica, Universidad  Cat\'olica de
Valpara\'{\i}so, Valpara\'{\i}so, Chile}
\maketitle
\begin{abstract}
A derivation of the effective lagrangian for non-relativistic quantum chromodynamics and the heavy quarks
effective field theory is given. Our calculation  provides of a simple and systematic method of calculation of the full
off shell effective lagrangian at tree level including all the $1/m$ corrections.
\end{abstract}

Presently QCD is the best candidate for describing the strong interactions of its  asymptotic freedom property at high
energies. However, in the infrared domain new additional problems  appear, namely, the effective coupling constant
increase for low energies and as a consequence phenomena such as confinement, hadronization or chiral symmetry
breaking  must studied using non-perturbatives techniques.

Non-relativistic quantum chromodynamics (NRQCD) and heavy quark effective field theory (HQET) are examples of
effective theories that describe approximately the low energy dynamics of QCD. Although both NRQCD
\cite{lepage} and HQET \cite{isgur} describe processes involving heavy quarks, they are physically different
theories. Indeed, while NRQCD describes processes at low transferred momentum, HQET only consider processes
where $k_0>>{\bf k}^2/2m$. From the physical point of view this difference is manifested in the structure of the
quarks propagators, in NRQCD
the propagator is
\begin{equation}
\frac{1}{k_0 - \frac{{\bf k}^2}{2m} + i\epsilon} \label{1}
\end{equation}
while in HQET is
\begin{equation}
\frac{1}{k_0 + i\epsilon} \label{2}
\end{equation}

However in spite of these physical differences, both theories are written in terms of an effective lagrangian coming
from QCD in the non-relativistic limit. Normally this effective lagrangian is derived using symmetry arguments
\cite{lepage1} and the coefficients of the operators involved in the expansion are obtained by using matching
conditions \cite{manohar}.

In this letter we would like to propose a new approach to the effective lagrangian calculation of NRQCD which has
the advantage that it allows a derivation of all the relativistic corrections of the (tree-level) effective lagrangian, and in
addition it is valid without using the equation of motion\cite{pepe}.

In order to discuss our results let us start considering the QCD lagrangian
\begin{equation}
{\cal L}_{QCD} = {\cal L}_G + {\cal L}_L + {\cal L}_H, \label{3}
\end{equation}
where ${\cal L}_G$ is the lagrangian for the gluons fields and  ${\cal L}_{L, H}$ is the fermionic part for the light
and heavy quarks respectively, {\it i.e.}
\begin{equation}
{\cal L}_{L,H} = {\bar \psi}_{L,H}\bigl[{iD \hspace{-.6em} \slash \hspace{.15em}} - m_{L,H} \bigr]\psi_{L,H},
\label{4}
\end{equation}
where ${D \hspace{-.6em} \slash \hspace{.15em}} = {\partial \hspace{-.6em} \slash \hspace{.15em}} + ig {A
\hspace{-.6em} \slash \hspace{.15em}}$.

In order to define the heavy quark mass one must neglect the hard gluons contributions while the
soft gluons one, by definition, are contained in the heavy quark mass.  This assumption is valid when the heavy quark
mass is much heavier than the scale of QCD and, under these conditions,  the heavy quarks can be considered as non-
relativistic particles. As a consequence of this, one focus the attention to the heavy modes sector of the partition
function
\begin{equation}
Z_H[A] = \int{\cal D}{\bar \psi}_H {\cal D}{\psi}_H \,e^{iS_H }. \label{5}
\end{equation}

Heavy quarks interact with the light modes through the gluon field and this is weak if measured at the scale of the
heavy  fermion mass  ( we will omit the subscript H from now on). In this case the original bispinor $\psi$ can be
written in terms of a slowly varying bispinor $\phi$ as\footnote{Notice that this reparametrizations of the fields can
be seen as changing the origin from where the energy $E$ is measured, {\it i.e.}, as defining $E_0 = E-m$ }
\begin{equation}
 \psi (x) = e^{-im t} \phi(x).  \label{6}
\end{equation}
where $\phi$ , in the leading approximation, carries no information about the heavy quark mass. The only
contribution coming from the mass of the heavy quarks appears in the corrections in powers of $1/m$.

Using (\ref{6}) one find that the heavy quark lagrangian is
\begin{equation}
{\cal L} = {\bar \phi}(i{D \hspace{-.6em} \slash \hspace{.15em}} - m(1- \gamma_0) )\phi. \label{7}
\end{equation}
This can be written explicitly in terms of the large $\varphi$ and small $\chi$ components of $\phi$ as

\begin{eqnarray}
{\cal L} &=& {\varphi}^{\dagger} iD{_0} \varphi + {\chi}^{\dagger} \bigl[i D{_0} + 2m\bigr] \chi +{\varphi}
^{\dagger}\,
i{\mathbf \sigma}\cdot {\bf D}  \chi \nonumber
\\
&&+{\chi}^{\dagger}\,i {\mathbf \sigma}\cdot {\bf  D} \varphi, \label{9}
\end{eqnarray}
where we have used the standard Dirac's representation for gamma matrices.
Thus, the partition function for the heavy quarks in terms of $\varphi$, and $\chi$, is
\begin{equation}
Z_H [A] = \int {\cal D} {\varphi}^{\dagger} {\cal D} \varphi {\cal D} {\chi}^{\dagger} {\cal D} {\chi}\,\, e^{iS_H}.
\label{10}
\end{equation}
The diagonalization of this lagrangian is straightforward. Indeed, if one performs the change of variables
(with unit Jacobian)
\begin{eqnarray}
\varphi^{'} &=& \varphi, \nonumber
\\
{\varphi}^{'\dagger} &=& {\varphi}^{\dagger}, \nonumber
\\
{\chi}^{'} &=& \chi + [ iD_0 + 2m]^{-1} \,i\,{{\mathbf \sigma}\cdot {\bf D}} \varphi, \label{11}
\\
{\chi}^{'\dagger} &=& {\chi}^{\dagger} +{\varphi}^{\dagger}\,i{{\mathbf \sigma}\cdot {\bf D}} \,[ iD_0 + 2m]
^{-1} \nonumber
\end{eqnarray}
in (\ref{5}), the new lagrangian reads (omitting the primes)
\begin{eqnarray}
{\cal L}&=&{\varphi}^{\dagger}\bigl[iD_0 + {{\mathbf \sigma}\cdot {\bf D}} \,{(iD_0 + 2m)}^{-1}\,{{\mathbf
\sigma}\cdot {\bf D}} \bigr] \,\varphi, \nonumber
\\
&+& \chi^{\dagger} \bigl[\, iD_0 + 2m \bigr] \chi. \label{13}
\end{eqnarray}
This lagrangian describes the (non-local) dynamics of relativistic heavy quarks in terms of  two components
spinors. One should note that $\varphi$ and $\chi$ appear decoupled and the integration in $\chi$ contributes with a
normalization factor (after choosing an appropriate gauge in this frame).

The non-relativistic limit of (\ref{13}) is straightforward because one expands the operator ${(iD_0 + 2m)}^{-1}$ in
powers of $1/m$, {\it i.e.}
\begin{equation}
{(iD_0 + 2m)}^{-1} = \frac{1}{2m} ( 1 - \frac{i}{2m}D_0 + \frac{i^2}{4m^2} (D_0)^2- ...).
\label{14}
\end{equation}
Then, the effective lagrangian for the heavy quarks coming from (\ref{13}) becomes
\begin{equation}
{\cal L} = {\cal L}^{(0)} + {\cal L}^{(1)} + {\cal L}^{(2)} + ....
\label{15}
\end{equation}

The first term in (\ref{15}) after to use $\sigma_i \sigma_j = \delta_{ij} + i \epsilon_{ijk}\sigma^k$,  reads
\begin{eqnarray}
{\cal L}^{(0)}={\varphi}^{\dagger}\,\bigl[ \,iD_0 + \frac{1}{2m}{\bf D}^2 + \frac{g}{2m} {\bf \sigma}\cdot{\bf
B}\,\bigr]
\varphi, \label{16}
\end{eqnarray}
which is just the lagrangian for a non-relativistic quark in a (chromo)magnetic field and it includes the Pauli term.

Even if expanded in this way, the resulting lagrangian is not written in the standard form, {\it i.e.} ${\cal L} = p {\dot
q} - {\cal H}$, from where one can infer directly the way the Hamiltonian looks like. To that end must use the
equations of motion for both ${\varphi}^{\dagger}$ and $\varphi$ when calculating those contributions higher that
${\cal L}^{(0)}$.

The higher order corrections are more laborious to find but the calculation is straightforward. Thus, the order
$1/m^2$ correction is
\begin{equation}
{\cal L}^{(1)}=\frac{g}{8m^2}{\varphi}^{\dagger}\bigl[ [{\bf D}, {\bf E}] + i{\bf \sigma}\cdot({\bf D}
\times {\bf E} - {\bf E}\times {\bf D})\bigr]\,\varphi, \label{17}
\end{equation}
where the terms in the RHS are the well known non-abelian Darwin and spin-orbit ones, and in the conmutator $[{\bf
D}, {\bf E}]$ an inner product is understood.

The next higher order terms ($1/m^3$) can be computed following the same procedure, thus one gets that
${\cal L}^{(2)}$ is
\begin{eqnarray}
{\cal L}^{(2)}&=&\frac{1}{8m^3}{\varphi}^{\dagger}\bigl[{\bf D}^4+
\, g \{{\bf D}^2, {\bf \sigma}\cdot{\bf B}\} + g^2 (\,{\bf B}^2 -{\bf E}^2)\nonumber
\\
&+& i\,g^2\,{\bf \sigma}\cdot({\bf B}\times{\bf B}-{\bf E}\times{\bf E})\bigr]
\,\varphi,
\label{18}
\end{eqnarray}
where the second line is a genuine QCD contribution.

The lagrangian  (\ref{15}) including $1/m^3$ corrections was written using dimensional and plausibility arguments
by Lepage {\it et. al.} in \cite{lepage1}. The above derivation  provides us of a systematic calculation method where
contact, spin and color terms simply does not exist in the non-relativistic limit. These last terms are discarded in the
Lepage {\it et. al.} analysis \cite{lepage1} by energetic considerations and, as a consequence, the approach proposed
here could be considered as a first principles calculation.

The equation (\ref{15}) is the NRQCD lagrangian in the rest frame, in an arbitrary frame (\ref{15}) becomes the
HQET lagrangian. For details see \cite{manohar}.

We finalize this paper making some comments concerning to the change of variables (\ref{11}). The new variables
diagonalize (\ref{9}) producing a non-local term in (\ref{13}). Although this non-local term  retains all the
information concerning to the relativistic corrections, one can see corrections such as Darwin, spin-orbit, etc. only
order by order. Thus, we could conjecture that the operator ${(iD_0 + 2m)}^{-1}$  contains all  the information
given by the  Foldy-Wouthuysen transformation \cite{balk}. In fact, this transformation is applied to the equation of
motion and then an extra condition is impossed in order to separate the small and large components of the spinor. In
our case, things are done in reverse order,{\it  i.e.}, a transformation is obtained that makes the separation between
small and large components and then the equations of motion are imposed.

After this paper was completed , we were informed about reference \cite{mannel} where a path integral derivation
of the effective action for HQET was done . In a rest frame our formula (\ref{13}) corresponds to the Mannel
{\it et. al. } effective action (eq. (20) in this reference) . However we would like to emphasize the differences between
both calculations; in our case instead of using projection operators as in \cite{mannel}, we propose the change of
variables (\ref{11}) obtaining (\ref{13}) straightforwardly. The explicit expansion of
${(iD_0 + 2m)}^{-1}$ permit to {\em derive} eq. (\ref{15})  ({\it i.e.} (14)-(16)) instead of conjecturing it as it has
been previously discussed.

This work was partially supported by grants 1970673, 1980788, 2990037 from Fondecyt -Chile and Dicyt-USACH.

\end{document}